\documentclass[preprint,tightenlines, aps]{revtex4}
\usepackage{graphicx}
\usepackage{dcolumn}
\usepackage{bm}
\begin{document}
\preprint{APS/123-QED}

\title{On the electromagnetic form factors of hadrons\\in the time-like region near threshold}
\author{O.D. Dalkarov}
\author{P.A. Khakhulin}
\author{A.Yu. Voronin}
\affiliation{Lebedev Physical Institute, Moscow, Russia}
\email{e-mail@gmail.com}
\date{\today}

\begin{abstract}
Hadron electromagnetic form factor in the time-like region at the boundary of the physical region is considered. The energy behavior of the form factor is shown to be determined by the strong hadron-antihadron interaction. Imaginary parts of the scattering lengths for $p\bar{p}$, $\Lambda\bar{\Lambda}$, $\Lambda\bar{\Sigma}^0 (\bar{\Lambda}{\Sigma}^0)$ and ${\Sigma}^0\bar{\Sigma}^0$ are estimated. Developed approach enables us to estimate imaginary part of the scattering volume from $D^*\bar{D^*}$ experimental data. The experiments to extract detailed information on the nearthreshold $B\bar{B}$ interaction from hadron form factor energy behavior are suggested.
\end{abstract}

\pacs{Valid PACS appear here}

\maketitle

\section{Introduction}
\label{intro}
The main goal of hadron electromagnetic form factors studies is to obtain  information about hadron structure. The most complete data on the behavior of the form factor as a function of four-momentum transferred is obtained for pion and nucleon. To investigate form factor of hadron $(h)$ two reactions are used: the reaction of elastic scattering of electrons by hadron $eh \to eh$ (so-called space-like region of the four-momentum transferred) and the reaction of hadron-antihadron pair production in the electron-positron annihilation $e^{+}e^{-} \to h\bar{h}$ or inverse reaction (time-like region).

High precision data on the electromagnetic form factor of a proton in the time-like region  became available due to PS--170 experiment  performed at LEAR (CERN) and recent BaBar experiment, where the reactions $p\bar{p} \to \Lambda\bar{\Lambda}, \Lambda\bar{\Sigma}^0, {\Sigma}^0\bar{\Sigma}^0$ were reported~\cite{1,2,3,4,5}.  BES collaboration~\cite{6} has observed the near-threshold enhancement of the $p\bar{p}$ - system in the $J/\psi \to Kp$ decay. These data demonstrate principally different behavior of the electromagnetic baryon form factor in the time-like region from that of a pion. The proton form factor drops  quickly  with momentum increasing from the threshold (approximately two times in the energy range from the threshold to $3.6 \:{\rm GeV}^2$, corresponding to change of c.m. relative momentum from zero to $100\:{\rm MeV}/c$).

To describe the behavior of the form factor different vector dominance models (VDM) are used. These models successfully reproduce the behavior of the pion form factor both in space-like and time-like regions~\cite{7} and the behavior of the nucleon (proton and neutron) form factors in the space-like regions~\cite{8}, but they fail in description of the proton form factor in the time-like region. Also several explanations of the energy behavior for proton form factor were considered after obtaining of the experimental data (see~\cite{9} and references therein).

Another approach was proposed~\cite{10,11,12,13} to understand physics of the behavior of the baryon electromagnetic form factor in the time-like region just near threshold. This approach takes into account final state interaction (interaction in the baryon-antibaryon system) as dominant physical reason giving  form factor energy behavior and its value near $B\bar{B}$ threshold. In this model the form factor is factorized  according to different physical processes. Form factor is presented as a product of a factor corresponding to singularities of transition  amplitude lying far from $B\bar{B}$ threshold and a factor reflecting strong final state interaction. The later gives the energy dependence of the form factor. Moreover, the behavior of the form factor appears to be directly connected to other observables in the $B\bar{B}$ system. For instance, it is possible to extract  imaginary part of baryon-antibaryon scattering length using the  form factor energy  dependence near $B\bar{B}$ threshold. Note that the consideration which was proposed in~\cite{10,11,12,13} is a natural consequence of the main features of quasi-nuclear model of low energy baryon-antibaryon interaction~\cite{14,15,16,17,18}. The near-threshold enhancement was predicted in~\cite{14,15,16,17,18} long time before first experimental indications.

Mentioned approach predicts peculiar behavior of the proton form factor in the time-like region and, moreover, it turns to be possible to predict the behavior of the neutron form factor. These predictions differ sufficiently from the predictions of standard VDM model. The neutron form factor can not exceed the proton one, whereas all vector dominance models obtain the neutron form factor five-ten times more than the proton one.

Established connection of the behavior of the hadron form factor with the interaction in the hadron-antihadron system gives us unique possibility to investigate these systems. The main advantage of the  baryon-antibaryon pair production in electron-positron collisions is that there is no initial state interaction and transition mechanism is well-determined. However, even existing facilities allows us to obtain such an information from the ''form-factor'' experiments with $e^{+}e^{-}$ beams (for instance for systems with hidden strangeness like $Y\bar{Y}$, or with hidden charm and beauty like $D\bar{D}$, $B\bar{B}$, etc.) In the following we will  discuss existing data on the $Y\bar{Y}$ production.

There is  additional group of experiments with $e^{+}e^{-}$ beams which can give more information about hadron-anti-hadron interaction in the final state. In case of nucleon-antinucleon these are experiments on  electron-positron annihilation into multipion systems which are dominant modes of nucleon-antinucleon annihilation. In this reactions  nucleon-antinucleon system appears as an intermediate state. The quantum numbers of the $N\bar{N}$ system are  those of photon, as far as $N\bar{N}$ are created from electron-positron pair through a photon. Even or odd amount of pions in the final state fixes isospin of a system. This gives unique opportunity to prepare the nucleon-antinucleon system in a state with definite quantum numbers. Let us mention that the deep-bump structure~\cite{19,20,21} observed in the electron-positron annihilation into six pions near nucleon-antinucleon threshold was explained in~\cite{11,12,13} as a property of $N\bar{N}$ intermediate state (namely the Green function zero of $\bar{N}$~\cite{22}). From these data the existence of a new narrow vector $N\bar{N}$ state was predicted~\cite{11,12,13}. In experiments with $\bar{p}$ scattering on nuclear targets the direct observation  of new subthreshold $N\bar{N}$ state is possible, but  requires special theoretical considerations due to cancelation of non-adiabatic and off-mass shell effects~\cite{23}.

The main advantage of the approach suggested in~\cite{10,11,12,13} and developed here is its ability to describe simultaneously the properties of nucleon-antinucleon interaction, electromagnetic form factor of the nucleon in the time-like region, and multipion electron-positron annihilation near $N\bar{N}$ threshold.

The article is organized the following way. Section 2 is devoted to the general properties of the form factor in the time-like region and the case of  nucleon form factor. In Section 3 we discuss form factor properties of other hadrons. Conclusion and proposals of new experiments are presented in Secton 4.

\section{General properties of the form factor}
\label{sec:1}
In this Section we investigate general properties of the hadron form factor in the time-like region near the boundary of the physical region. For different hadrons we can have different number of form factors, depending on hadron spin. We will formulate the main results for a proton and neutron, the generalization  for other hadrons will be done in Section 4.

The form factor of the nucleon ($N$) in the time-like region is determined from the reaction of $e^{+}e^{-}$-annihilation into nucleon-antinucleon pair $e^{+}e^{-} \to N\bar{N}$ or vice versa. This is so-called s--channel in contrast to t--channel corresponding to $eN \to eN$ scattering or to the nucleon form factor in the space-like region.

The differential cross section $d\sigma/d\Omega$ of the reaction $p\bar{p} \to e^{+}e^{-}$ is connected with the form factor of the proton in the vicinity of $p\bar{p}$ threshold by the formula
\begin{equation}
	\frac{d\sigma}{d\Omega} = \frac{\alpha^2}{32kE}\left(|G_{M}|^2 (1+\cos^2\theta) + \frac{{4M_p}^2}{s}|G_E|^2 \sin^2\theta\right),
\end{equation}
here $k$ and $E$ is center-of-mass momentum and energy in the $p\bar{p}$ system, $\theta$ is angle in c.m.s., $\alpha$ is the fine structure constant, $M_p$ is the proton mass. $G_E$ and $G_M$ are electric and magnetic form factors of the proton correspondingly. They are connected to Pauli form factors $F_1$ and $F_2$:
\begin{equation}
	G_M = F_1+F_2, \;\;\; G_E = F_1-\frac{q^2}{4{M_N}^2}F_2,
\end{equation}
here $q$ is four-momentum transferred, which is equal to $t$ in the center-of-mass system of the reaction $eN \to eN$ and to $s$ in the reaction $e^{+}e^{-} \to N\bar{N}$, threshold of the latter reaction corresponds to $q^2 = -4{M_N}^2$ ($M_N$ is the nucleon mass).

At the $p\bar{p}$ threshold $G_E$ and $G_M$ are equal and for simplicity hereafter they are taken to be equal in the kinetic energy region of few tens of MeV near the threshold.

\begin{figure}[h]
\includegraphics{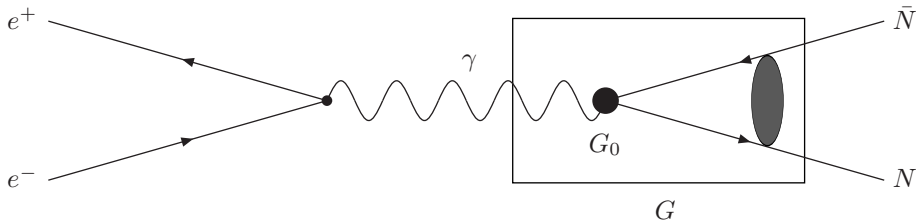}
\caption{The diagram corresponding to electron-positron annihilation into the nucleon-antinucleon system.}
\label{diagram}
\end{figure}

Before doing any calculations we can make some conclusions about nucleon form factor near threshold. Let's consider a diagram corresponding to the process $e^{+}e^{-} \to N\bar{N}$ (Fig.~\ref{diagram}). Grey block in this diagram presents the final state interaction in the system $N\bar{N}$. We know that this interaction is very strong and even can produce bound states in the $N\bar{N}$ system~\cite{14,15,16,17,18}. Left part of the diagram corresponds to the transition amplitude from $e^{+}e^{-}$ pair into $N\bar{N}$ pair. Black circle in this transition amplitude denotes a connection between a photon and $N\bar{N}$ pair, which can be realized for example by vector mesons ($\rho$ or $\omega$).

%

Since $N\bar{N}$ annihilation interaction is short-range the form factor  can be factorized as follows:
\begin{equation}
\label{ff}
	G = G_0|A_l(k)|.
\end{equation}
Here $A_l(k)$ is connected to the amplitude of the normalized wave-function near zero:
\begin{equation}
\Psi_l(r) \sim A_l(k) (kr)^{l}.
\end{equation}

 The factor $G_0$ corresponds to singularities far from $N\bar{N}$ threshold (for instance, to a connection of the photon to the nucleon through $\rho$- or $\omega$-exchanges, as it can be written in usual vector dominance models) and is practically constant  in the kinetic energy region of few tens of MeV in the $N\bar{N}$ center of mass system. The factor $A_l(k)$ reflects the influence of strong final state interaction and contains main energy dependence of the form factor (\ref{ff}) in the near-threshold region.
The general form of energy behavior is given by the following expression (see Appendix A) :
\begin{equation}\label {Gform}
G=G_0\frac{\exp(-{\rm Im}\delta(k))}{\tau(k^2)}
\end{equation}
Here $\delta(k)$ is the $N\bar{N}$ scattering phase, $\tau(k^2)$ is an function of $k^2$, determined by the properties of  the $N\bar{N}$ potential.
In the case of $s$--scattering $A_0(k) \propto \Psi({\bf 0})$ and eq.~(\ref{ff}) becomes identical to the one obtained in~\cite{10}.
In the region just near threshold $G(k)$  (\ref{Gform}) has the form (see Appendix):
\begin{equation}\label{Gthresh}
	G = G_0\frac{\exp({\rm Im}\alpha_0 k)}{\tau(0)}\approx{\rm const} (1+{\rm Im}\:\alpha_0\:k),
\end{equation}
where $\alpha_0$ is the $N\bar{N}$ $s$--wave scattering length.
 The appearance of  a linear in $k$ term is a direct manifestation of the threshold behavior and the final state \emph{inelastic} interaction.

For very small $k$ the role of the Coulomb corrections is of special interest (see Appendix A). In this case we get for the factor $A_0(k)$ in  Eq.(\ref{ff}) :
\begin{equation}
	|A_0(k)| = {\rm const}\: C \:(1+C^2{\rm Im}\alpha_0 k).
\end{equation}
Here $C$ is the Gamow factor
\begin{equation}
	C^2 = \frac{2\pi/k|a_c|}{1-\exp(-2\pi/k|a_c|)}.
\end{equation}
and $|a_c| = 2\hbar^2/(M\:e^2)$ is the Coulomb length unit, $M$ is $N\bar{N}$ reduced mass. The Coulomb corrections becomes important when $2\pi/k|a_c| > 1$ , for $p\bar{p}$ system it means $k < 20\: {\rm MeV}/c$.

In the case of $p$--wave the $A_1(k)$ factor near threshold is given by the following expression (Appendix A):
\begin{equation}\label{Gpwave}
	|A_1(k)| ={\rm const}\: \frac{\exp({\rm Im} \alpha_1 k^3)}{\tau(k^2)}\simeq {\rm const}\:(1 + {\rm Re}c\: k^2 +{\rm Im}\alpha_1 k^3)
\end{equation}
Here $\alpha_1$ is the scattering volume, $r$ is $k$--independent coefficient.

In the region not too close to the threshold ($k\alpha\sim 1$), it is not possible  to get  model independent information about the properties of $N\bar{N}$ potential. For such a  purpose  we suggest a simple phenomenological model of $N\bar{N}$ interaction based on the following assumptions. First, the tail of $N\bar{N}$ potential is considered  attractive and above a certain matching interbaryonic distance $R_c$ it can be approximated by exponential potential:
 \begin{equation}\label{PotMod}
 V_{N\bar{N}}(r>R_c)=-U_0 \exp(-(r-R_c)/\rho)
 \end{equation}
 here $\rho$ is a free diffuseness parameter.

 Second, the effective depth of $N\bar{N}$ interaction below $R_c$ is much greater than a collision energy of interest. It means that at matching distance $R_c$ the $N\bar{N}$ non-relativistic wave-function $\Psi(R)$ obeys the  \emph{energy independent} boundary condition, which can be conveniently  parameterized in the following way:
 \begin{equation}\label{BoundMod}
 (r\Psi)'/(r\Psi)|_{r=R_c}=\frac{z(R_c)}{2\rho} \cot (\Phi_0)
 \end{equation}
 Here $z(R)=2\rho \sqrt{2 M U_0}\exp(-(R-R_c)/(2\rho))$, $\Phi_0$ is a free  parameter.

 We use the closed form solution of the Schr\"{o}edinger equation with the exponential potential (see Appendix B) to get the following expression for the  $s$--wave form-factor:
 \begin{equation}\label{fmodel}
 |A_0(k)|={\rm const}\sqrt{\frac{\sinh(2 \pi k \rho)}{2 \pi k \rho}}\frac{1}{|\cos(\Phi+i \pi k \rho )|}
 \end{equation}
Here we introduced phase parameter $\Phi\equiv\Phi_0+z(R_c)-\pi/4$.
 The  above expression is justified when $z(R_c)\gg 1$ and $ |U_0|\gg k^2/(2 M)$.

 Let us consider the main properties of the above introduced model. We have two model parameters: the diffuseness $\rho$ of the potential tail and the complex  phase $\Phi$, which describes the influence of the deep inner part of $N\bar{N}$ interaction. In the limit of small $k$ we return  to  Eq.(\ref{Gthresh}) with the  imaginary part of the scattering length given by the following expression:
 \begin{equation}\label{Ima}
{\rm Im}\alpha_0 = -\pi \rho\:{\rm Im}\tan(\Phi)
\end{equation}
In case   ${\rm Im}\:\Phi \gg 1$ (strong absorption by the inner part of $N\bar{N}$ interaction) the form factor becomes insensitive to parameter $\Phi$. In such a case the imaginary part of the scattering length turns to be :
\[{\rm Im}\alpha_0=-\pi \rho\]
while the form factor is:
\[|A_0(k)|={\rm const}\sqrt{\frac{\sinh(2 \pi k \rho)}{ \pi k \rho}}\exp(-\pi k \rho)\]
In the opposite case of weak absorption ${\rm Im}\:\Phi\ll 1$ the form factor is sensitive to the phase $\Phi$, especially when ${\rm Re} \Phi \rightarrow \pi/2$. This corresponds to the appearance of a quasi-bound $N\bar{N}$ state close to the threshold. In such a resonant case the value of the  imaginary part of the scattering length can be much greater than the diffuseness of potential tail:
\[{\rm Im}\alpha_0\approx-\pi \rho\: \frac{1}{{\rm Im}\Phi}\]
The form factor in the resonant case turns to be:
\[|A_0(k)|={\rm const}\sqrt{\frac{\sinh(2 \pi k \rho)}{2 \pi k \rho}}\frac{1}{\sinh({\rm Im} \Phi+ \pi k \rho) }\]
The important property of suggested model is the ability to describe the form factor energy behavior beyond the scattering length  and effective range approximations i.e. for $k\geq 1/\rho$. In the limit $1 \ll k \rho \ll z(R_c)$ the form factor behaves like:
\[|A_0(k)|={\rm const}\frac{1}{\sqrt{ \pi k \rho}} \]
So far the above model  describes the transition from fast exponential decay $\exp({\rm Im} \alpha_0 k)$ of the form factor just near the threshold to the  $1/\sqrt{\pi k \rho}$ behavior away from the threshold. Such a transition takes place when $k \pi \rho \sim 1$. The experimental data fit in such a transition region enables us to extract diffuseness parameter $\rho$, while the data fit  near the threshold gives the imaginary part of the scattering length and so far the phase parameter $\Phi$. This property of the model gives an opportunity to distinguish the nearthreshold resonance in the form factor behavior, characterized by the imaginary part of the scattering length much larger than the potential tail diffuseness $|{\rm Im}\alpha_0|\gg \pi \rho$.

\begin{figure}[ht]
\includegraphics{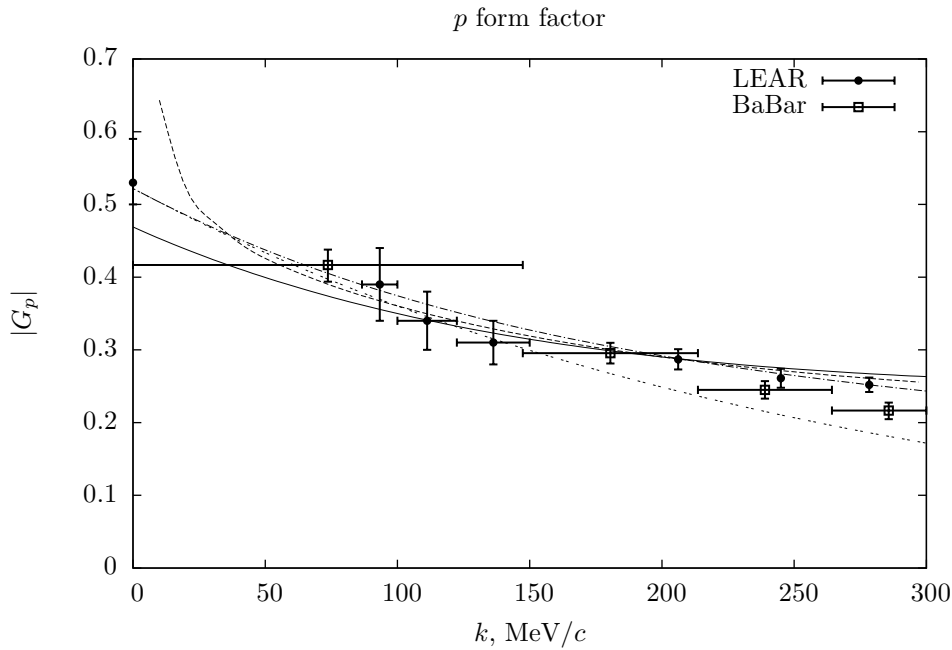}
\caption{Proton electromagnetic form factor in the time-like region. Experimental data are taken from~\cite{1,3}. Solid and dotted lines are calculations using optical model~\cite{26,27} with and without Coulomb corrections. Dash-dotted line represents the data fit by the model Eq.(\ref{fmodel}). Double-dotted curves are the extrapolation of nearthreshold form factor behavior by means of exponential curves $\sim \exp({\rm Im}\alpha_0 k)$.}
\label{protonff}
\end{figure}

We demonstrate the applicability of this model to the case of $p\bar{p}$ $s$--wave form factor. Precise experimental data on the proton form factor near the threshold were obtained in the LEAR and BaBar experiment~\cite{3}. These data  are presented in Fig.~\ref{protonff}. We note the sharp peak at small relative momenta with decreasing of the form factor in two times. We perform comparison  between the numerical solution of the Schr\"{o}edinger equation with the OBE potential \cite{26,27}, the model form factor given by Eq.(\ref{fmodel}) and its nearthreshold form Eq. (\ref{Gthresh}).




The $p\bar{p}$ scattering length extracted from the fit of  LEAR experimental data using Eq.(\ref{fmodel}) is ${\rm Im}\alpha = -(0.73\pm0.05)\:{\rm fm}$. 

This value is in good agreement with the imaginary part of the scattering length value (${\rm Im}\:\alpha({}^3 S_1) = -0.8\:{\rm fm}$) obtained from the data on the protonium ($p\bar{p}$ Coulomb atom) triplet $s$--levels shifts and widths~\cite{25}.

One can see that the model form factor Eq.(\ref{fmodel}) reproduces the  experimental data up to $k=400 $ MeV/$c$, where both the scattering length and effective range approximation fails. We note that  realistic OBE $p\bar{p}$ potential is a superposition of Yukawa-type potentials with different diffuseness and effective depth. However the potential tail, important for the form factor energy behavior, can be satisfactory  approximated by the "averaged" exponential tail. The model parameters for $p\bar{p}$ were found to be $\rho=0.3\:{\rm fm}$ and $\Phi=0.01+i1.02$

By using  experimental data on the proton form factor we can predict a value of the neutron form factor  near the threshold. The definition of the nucleon form factor in terms of the isoscalar (isospin is equal to $0$) $G(0)$ and isovector $G(1)$ (isospin 1) form factors is:
\begin{equation}
	G_p = |G(0)+G(1)|,\: G_n = |G(0)-G(1)|.
\end{equation}

We can use this decomposition and express $G_p$ and $G_n$ in terms of input form factors $G_0$ and wave functions of final state in pure isospin states:
\begin{equation}
	G_p = |G_0(0)\Psi_0({\bf 0}) + G_0(1)\Psi_1({\bf 0})|,
\end{equation}
\begin{equation}
	G_n = |G_0(0)\Psi_0({\bf 0}) - G_0(1)\Psi_1({\bf 0})|.
\end{equation}

Depending on the relative values of $G_0(0)$ and $G_0(1)$ there are two possible cases.

\begin{figure}[ht]
\includegraphics{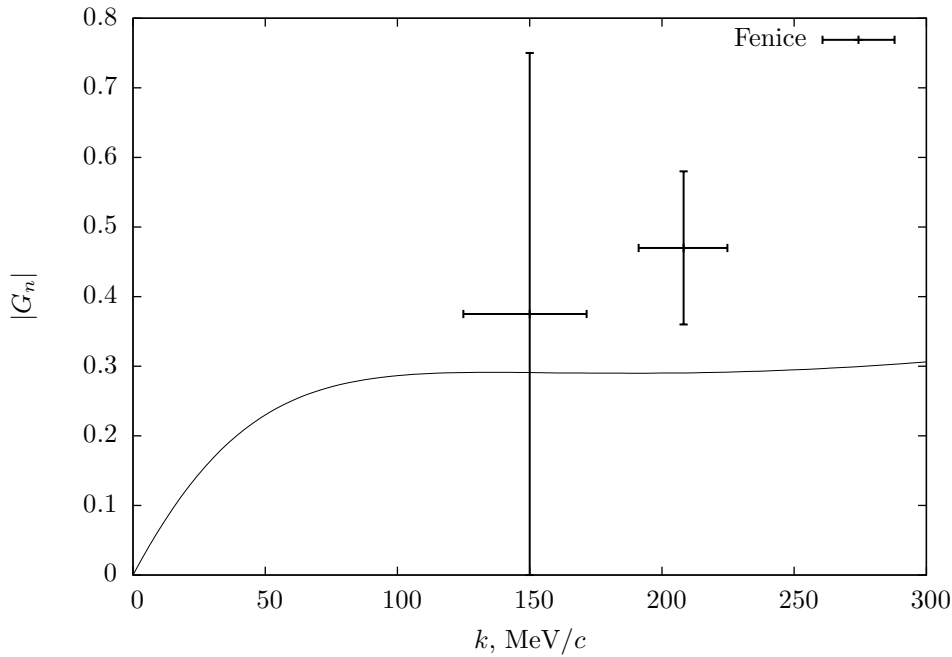}
\caption{Neutron form factor in the time-like region. Experimental data from~\cite{19}. Solid line is our calculation using optical model~\cite{26,27} with normalization at $k = 200\:{\rm MeV}/c$.}
\label{neutronff}
\end{figure}

First, $G_0(0) \approx G_0(1)$, i.e. there is no sufficient difference between isoscalar and isovector input form factors. In this case in the previous formulas we can take out of brackets the common factor $G_0$. We see immediately that neutron form factor $G_n$ is equal to or less than proton one. Note that in this case the energy behavior of the neutron form factor can be different from the proton one. Just near threshold it can be practically constant or even decreasing function of the energy. The calculations of the neutron form factor using optical model~\cite{26,27} in this case are presented in Fig.~\ref{neutronff}. Note that the close predictions were done in coupled channels model in ref.~\cite{11,12,13}. Preliminary experimental data~\cite{19,20,21} indicates on realization of this possibility.

Second, one of the input form factors is dominant, for instance, $G_0(1) \gg G_0(0)$. This is possible if  the form factors are determined by $\rho$- and $\omega$- mesons correspondingly. In this case   $G_0(1)$ dominates because $\rho$-meson has a product of coupling constants with nucleon and photon larger than that for $\omega$-meson. So one can neglect $G_0(0)$ contributions to the nucleon form factor and  get that proton and neutron form factors are approximately equal.

Therefore, in our model the neutron form factor does not exceed proton one in any case. We note that  VDM models  give nonrealistic neutron form factor, which turns to be five or even ten times larger than that of proton.

\section{Form factors of other hadrons.}
\label{sec:4}
The consideration presented above can be directly applied to investigation of the form factor of any hadron in time-like region.

\begin{figure}
\includegraphics{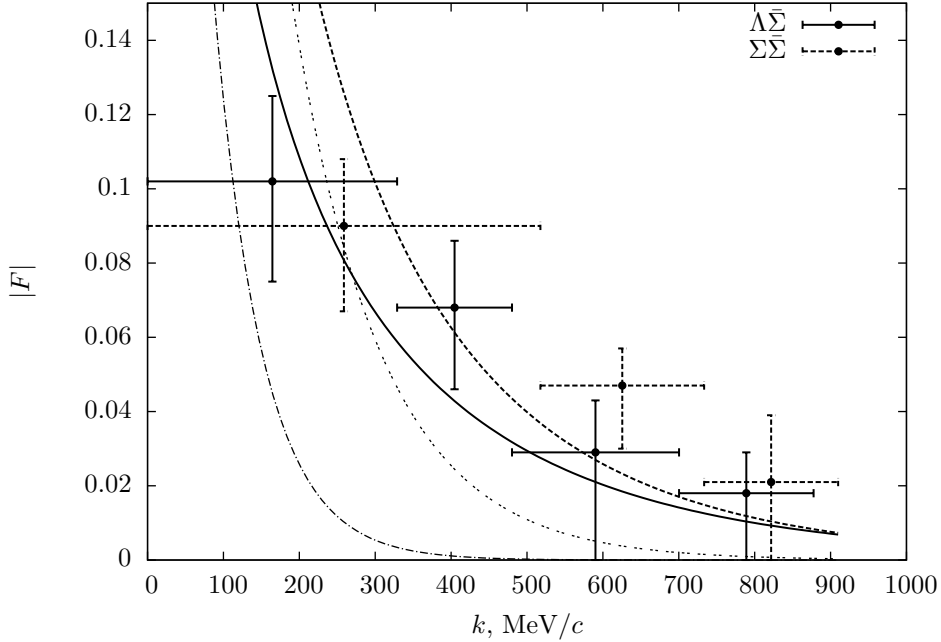}
\caption{Baryon effective form factors. Experimental data are taken from BaBar experiment~\cite{5}. Solid and dotted curves are $\Lambda\bar{\Sigma}^0 (\bar{\Lambda}\Sigma^0)$ and $\Sigma^0\bar{\Sigma}^0 \;$ experimental data fits by Eq.(\ref{fmodel}) correspondingly. Dash-dotted curves are the extrapolation of nearthreshold form factor behavior by means of exponential curves $\sim \exp({\rm Im}\alpha_0 k)$ .}
\label{baryef}
\end{figure}

New data on $\Lambda \bar{\Lambda}$, $\Sigma^0 \bar{\Sigma}^0$ and $\Lambda \bar{\Sigma}^0$ form factors became available in experiments of BaBar group. In~\cite{5} the effective form factor is introduced:
\begin{equation}
	{|F(m)|}^2 = \frac{2\tau {|G_M(m)|}^2+{|G_E(m)|}^2}{2\tau + 1},
\end{equation}
where $\tau = m^2/4{m_B}^2$, $m$ is mass of hadronic system and $m_B$ is baryon mass. In the absence of precise data we will assume that $|F|=|G|$. This assumption simplifies the  scattering lengths  imaginary parts estimation. Experimental data for $\Lambda\bar{\Lambda}$ and $\Lambda\bar{\Sigma}$ and their typical fits by expression (\ref{fmodel}) as well as by exponential curves $|F| \approx C\exp({\rm Im}\:\alpha\:k)$ are presented  in Fig.~\ref{baryef}.


The extracted scattering lengths values are presented in Table~\ref{scatlen}.

\begin{table}
\caption{Scattering lengths in $B\bar{B}$ system.}
\label{scatlen}
\begin{center}
\begin{ruledtabular}
\begin{tabular}{cccc}
\hline\noalign{\smallskip}
 &  ${\rm Im}\:\alpha, {\rm fm}$ & $\rho, {\rm fm}$ & ${\rm Im}\:\Phi$\\
\noalign{\smallskip}\hline
$\Lambda\bar{\Lambda}$ & $-1.3 \pm 0.6$ & $0.4 \pm 0.3$ & $0.8\pm 0.3$ \\
\noalign{\smallskip}\hline
$\Lambda\bar{\Sigma}^0 (\bar{\Lambda}\Sigma^0)$ & $-2.4\pm 0.7$ & $0.7\pm 0.1$ & $1.02 \pm 0.02$ \\
\noalign{\smallskip}\hline
$\Sigma^0\bar{\Sigma}^0$ & $-2.5\pm 0.8$ & $0.6\pm 0.1$ & $0.8 \pm 0.2$ \\
\noalign{\smallskip}\hline
\end{tabular}
\end{ruledtabular}
\end{center}
\end{table}


The closest threshold to the $N\bar{N}$ one is a threshold of $\Lambda \bar{\Lambda}$ production in $e^{+}e^{-}$-annihilation.

Let's estimate the value of the lambda form factor near $\Lambda \bar{\Lambda}$ threshold under the following assumptions. We consider the $\Lambda \bar{\Lambda}$ pair production mechanism via a vector meson (with dominant contribution of $\rho$-meson). The value of $G_0$ for lambda is proportional to the coupling constants of $\rho$ meson with photon and $\rho$-meson with lambda. The former is known from the experiment. The latter can be estimated from the SU(3)--relations. Both are of the same order as for a nucleon. Final state interaction in $\Lambda \bar{\Lambda}$ system (in pure isospin ($I=0$) state) according to the existing approaches is approximately the same as in case of $N\bar{N}$. So we expect lambda and nucleon form factors  have the same order of magnitude.

\begin{figure}[ht]
\includegraphics{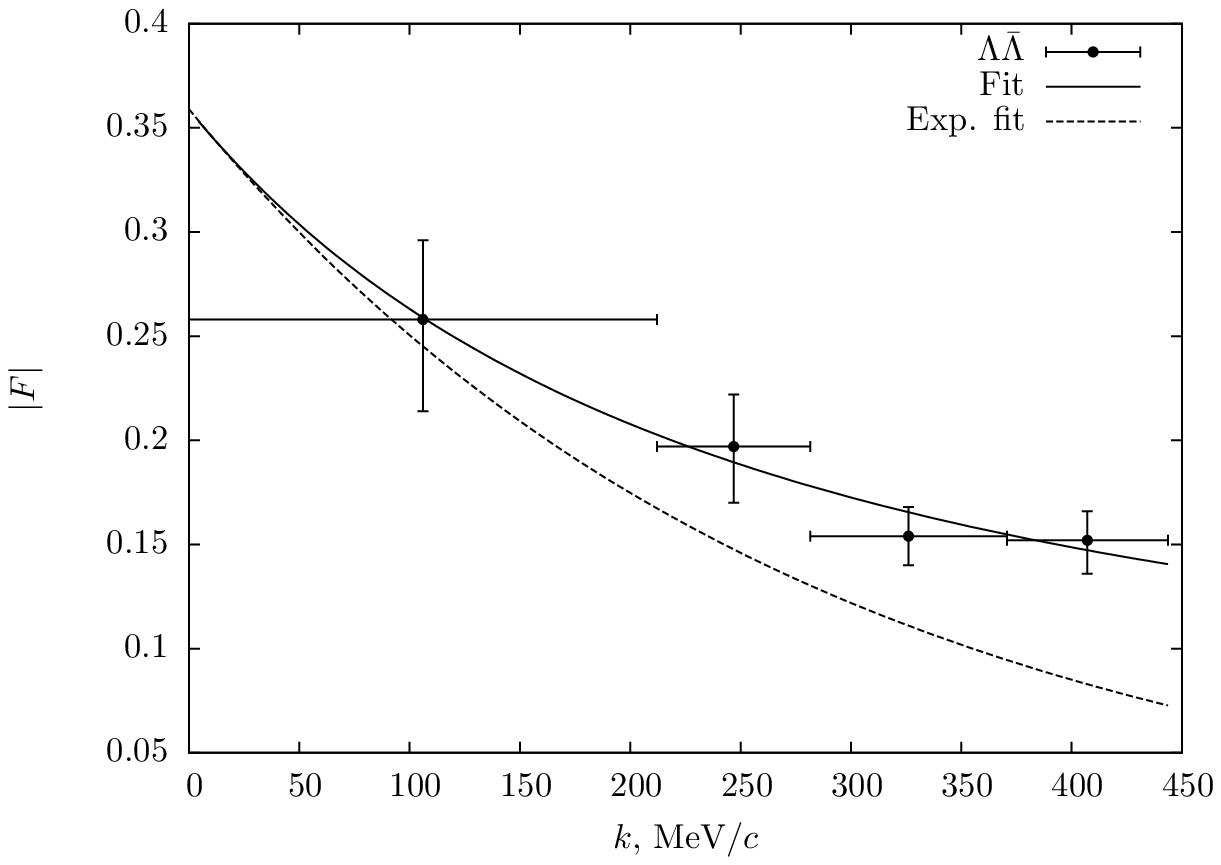}
\caption{Experimental points for $\Lambda\bar{\Lambda}$ system are taken from~\cite{5}. Solid line corresponds to the  fit by model Eq.(\ref{fmodel}), dotted line corresponds to the exponential extrapolation of the nearthreshold form factor behavior $\sim \exp({\rm Im}\alpha_0 k)$  }
\label{fit_LL}
\end{figure}

 The present data (though insufficient) indicate  possible nearthreshold $\Lambda\bar{\Lambda}$ resonance. On Fig.~\ref{fit_LL} we present a fit with parameters $\rho=0.1\:{\rm fm}$ and $\Phi=1.57+i0.55$, ${\rm Im}\:\alpha = -0.71\:{\rm fm}$. The real part of the phase shift $\Phi$ is found rather close to $\pi/2$, which is the indication of the $\Lambda\bar{\Lambda}$ nearthreshold state. More precise experimental data are required for determination of the scattering properties with much higher significance.

It follows from our consideration that fast decrease of the form factor with increasing momentum from the threshold  is the consequence of the absorption in the final state interaction. Let us mention, that such a decay can not be seen in the pion form factor, because $\pi^{+}\pi^{-}$ system has only elastic scattering and has no absorbtion at the threshold.

Let us turn to the case of the $D^*\bar{D^*}$ system. Experimental data became available due to experiments by CLEO-c. In the following we use the $D^*\bar{D^*}$ production cross-section parametrization  suggested in~\cite{28}:
\begin{equation}
	\sigma = \frac{14\pi\alpha_{fs}^2}{3m^2} v_{D^*}^3 R_4
\end{equation}
The value $R_4$  is proportional to the form factor squared.

\begin{figure}
\includegraphics{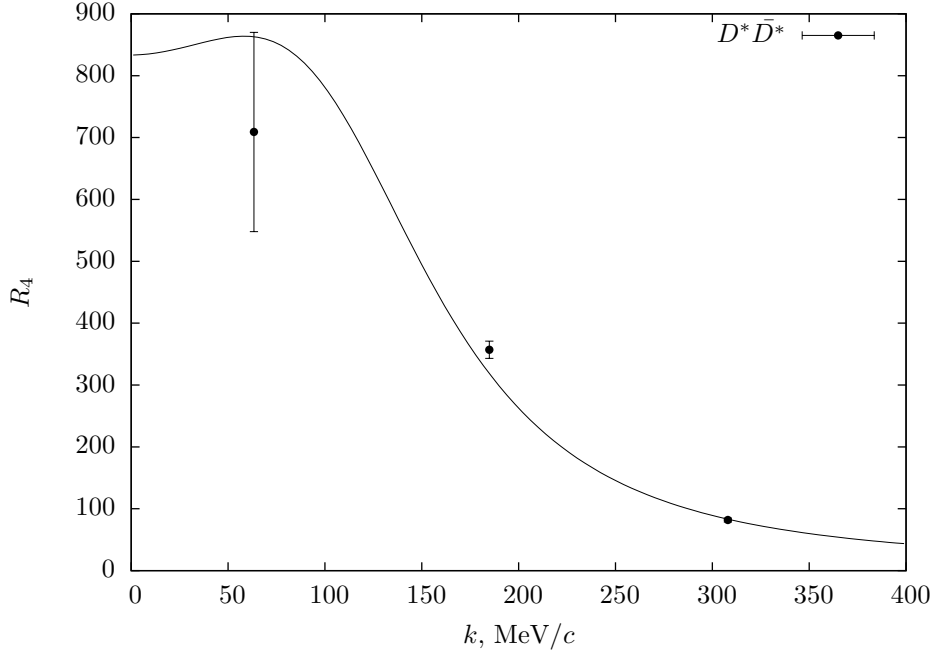}
\caption{Function $R_4$ \cite{28} fit by numerical solution of the Schr\"{o}dinger equation within model Eq.(\ref{PotMod},\ref{BoundMod}).}
\label{R4}
\end{figure}

 $D^*\bar{D^*}$ mesons are produced in $p$--wave. We apply the  model Eq.(\ref{PotMod},\ref{BoundMod}) and obtain the corresponding form-factor numerically.
Results are presented in Fig.~\ref{R4}. We have found a fit to the experimental data with $\rho = 0.3\:{\rm fm}$, ${\rm Im}\:\alpha_1 = -0.61\:{\rm fm^3}$.
 In spite of rather good agreement between theoretical fit and experimental data the obtained value of imaginary part of scattering volume should be treated only as an estimation because of the lack of CLEO-c data in the nearthreshold region.

\section{Conclusion}
\label{sec:5}
The analysis of the experimental data shows that the behavior of the electromagnetic form factor of a hadron is mostly determined by the interaction of the hadron-antihadron in the final state. Therefore the measurements of the form factor properties can serve as a  fruitful source of information about hadron-antihadron interaction, especially in situation when direct investigation of this interaction is unavailable.

To obtain more elaborate information about hadron-antihadron interaction the following experiments could be suggested:
\begin{itemize}
\item Precise measurements of the proton and neutron form factors in the time-like region just near threshold of the reaction $e^{+}e^{-} \to N\bar{N}$ give us opportunity of high quality determination of $N\bar{N}$ scattering parameters.
\item Further investigation of the strange and charm particles electromagnetic form factors in the time-like region in non-relativistic region of relative momentum less than few hundreds of ${\rm MeV}/c$.
\item There is a possibility to discover a ${}^3 S_1{}$ quasinuclear bound states in the $B\bar{B}$ system, which can manifest themselves as a heavy vector meson. To do it, the experiment to measure proton form factor near $B\bar{B}$ threshold is desirable, because these states will manifest themselves as the bumps in the form factor behavior ($e^{+}e^{-} \to B\bar{B} \to p\bar{p}$).
\item Bound states with photon quantum numbers in hadron-antihadron systems will manifest themselves also as a deep-bump structure in the electron-positron transition into main annihilation channels of these systems. In particular it will be interesting to search for phenomena connected with such a vector meson state in the $\Lambda\bar{\Lambda}$  system near threshold in the reaction $e^{+}e^{-} \to K\bar{K}4\pi$ by the analogy with deep-bump structure in 6$\pi$ annihilation channel near $N\bar{N}$ threshold. Note that the existence of quasinuclear ${}^3S_1$ vector state just near $\Lambda\bar{\Lambda}$ threshold was considered in~\cite{29}.
\item To determine interaction in the systems with hidden new quantum numbers, the experiments on precise measurement of the cross sections $e^{+}e^{-} \to K\bar{K}$, $D\bar{D}$, $F\bar{F}$, $B^{+}\bar{B}^{+}$, $B^{-}\bar{B}^{-}$ etc. near corresponding thresholds can be very informative.
\end{itemize}

\appendix
\section{}
\label{appndx}
We present here the derivation of the  form factor in terms of the Jost function. The form factor is proportional to the $A_l(k)$ that is defined from the following relation
\begin{equation}
	\psi_l(r; k) = A_l(k) \phi_l(r; k),
\label{def}
\end{equation}
where $\psi_l$ and $\phi_l$ are solutions of the Shr\"{o}dinger equation such that
\begin{equation}
	\psi_l(r; k) \sim \frac{i}{2} \left(\hat{h_l}^{-}(kr) - S_l(k) \hat{h_l}^{+}(kr)) \right),
\end{equation}
as $r \to \infty$,
\begin{equation}
	\phi_l(r; k) \sim \hat{j_l}(kr),
\end{equation}
as $r \to 0$.  Here $S_l$ is the partial S--matrix element. Functions included in these formulas are Riccati--Bessel and Riccati--Hankel functions:
$$
	\hat{j_l}(z) \equiv \sqrt{\frac{\pi z}{2}} J_{l+1/2}(z),
$$
$$
	\hat{n_l}(z) \equiv (-1)^l \sqrt{\frac{\pi z}{2}} J_{-l-1/2}(z),
$$
$$
	\hat{h_l}^{\pm} \equiv \hat{n}_l(z) \pm i \hat{j}_l(z).
$$
By the definition of the Jost function $f_l(k)$~\cite{30} we see that:
\begin{equation}
	A_l(k) = 1/f_l(k).
\label{A_l}
\end{equation}

Let's investigate form factor behavior at small $k$. According to integral representation of the Jost function~\cite{30}:
\begin{equation}
	f_l(k)  = 1+\frac1{k} \int \hat{n_l} U \phi_l dr + \frac{i}{k} \int \hat{j_l} U \phi_l dr
\label{intf}
\end{equation}
where $U(r) \equiv 2mV(r)$; and for the partial amplitude $F_l(k)$:
\begin{equation}
	F_l(k) = - \frac{1}{k^2 f_l(k)} \int \hat{j_l} U \phi_l  dr.
\label{partamp}
\end{equation}
Functions included in these formulas has the form
$$
	\hat{n_l} \propto \frac{1}{k^l} \times \left( a_0 + a_2 k^2 + ... \right),
$$
$$
	\hat{j_l} \propto k^{l+1} \times \left(b_0 + b_2 k^2 + ... \right),
$$
$$
	\phi_l \propto k^{l+1} \times \left(c_0 + c_2 k^2 + ... \right)
$$
with $a_0, a_1, ..., b_0, b_1, ... c_0, c_1, ...$ independent on $k$. Substituting these expansion forms in~(\ref{intf}) we get
\begin{equation}
	f_l(k) = h_0 + h_2 k^2 + ... + ig k^{2l+1}+ ... .
\end{equation}
To connect coeffitients $h_0, h_2, ..., g, ...$ with scattering parameters we use equation~(\ref{partamp}) to obtain
\begin{equation}
	F_l(k) = -\frac{g k^{2l}+ ...}{h_0 + h_2 k^2 + ... + ig k^{2l+1}}.
\end{equation}
When $k$ is small enough it is useful to apply scattering length approximation $F_l(k) \approx -\alpha_l k^{2l}$. Comparing the two above expressions we get that $g/h_0 = \alpha_l$. So, we have established connection between the  scattering length and the form factor for small $k$.

Particulary for $l=0$
\begin{equation}
	A_0(k) \approx {\rm const}\:(1-i\alpha_0 k),
\label{A0}
\end{equation}
for $l=1$
\begin{equation}
	A_1(k) \approx {\rm const}\:(1 + c k^2-i\alpha_1 k^3).
\end{equation}



Now let's find the form factor in the presence of the Coulomb force. We will investigate the most important case of $l=0$.
Now instead of equation~(\ref{intf}) we have
$$
	f(-k) = f^{c}(-k) + \frac{1}{k} \int_0^{\infty} dr H^{+}(r) U(r)\phi(r)
$$
where $U(r)$ is the effective strong force complex potential which does not include Coulomb interaction, $H^+$ is the Coulomb outgoing solution:
\begin{equation}
	H^{\pm} \to \exp \left( \pm i \left( kr - \frac{1}{ka_c} \ln 2kr \right) \right), \qquad r \to \infty.
\end{equation}
and $$
	f^{c}(-k) = \frac{\exp(-i\sigma_0)}{C}.
$$
Here
$$
	C^2 = \frac{2\pi/ka_c}{\exp(2\pi/ka_c)-1}
$$
is the Gamow factor, $a_c = \hbar^2/m Z_1 Z_2 e^2$ is the Coulomb length and $\sigma_0 = {\rm arg}\Gamma \left(1+i/ka_c\right)$ is the Coulomb phase.
Then it follows that
$$
	f(-k) = f^{c}(-k) + \frac{1}{Ck}\exp(-i\sigma_0) \int_0^{\infty} dr N(r) U(r)\phi(r) + \frac{iC}{k}\exp(-i\sigma_0) \int_0^{\infty} dr J(r)U(r)\phi(r).
$$
Here $J(r)$ is the regular Coulomb solution:
\begin{equation}
	J(r) = \exp(ikr) kr {}_1 {\rm F_1} \left(1+\frac{i}{ka_c}, 2, -2ikr\right),
\end{equation}
The first integral is proportional to $1+O(k^2)$ and the second one is proportional to $k\times(1+O(k^2))$. We will denote the first integral as $h$ and the second one as $g$.

The integral representation for the scattering amplitude has the form :
$$
	F = F^{c} - \frac{1}{k^2} \frac{1}{f(-k)}\frac{1}{f^{c}(-k)} \int_0^{\infty} dr \phi(r) U(r) J(r),
$$
where $F^{c} = (\exp(2i\sigma_0)-1)/2ik$ - coulomb scattering amplitude. In the above expression we have replaced the radial wave function and the pure Coulomb radial wave function with their representations in terms of regular solutions and the Jost functions. We get so far:
\begin{equation}
	F(k) = F^{c}(k) - \frac{\exp(2i\sigma_0)}{k^2} \frac{\int_0^{\infty} dr \phi U J}{\frac{1}{C^2} + \frac{1}{C^2 k}\int_0^{\infty} dr N U \phi + \frac{i}{k} \int_0^{\infty} dr J U \phi}
\end{equation}
or equally
\begin{equation}
	F(k) = F^{c}(k) - \exp(2i\sigma_0) \frac{g}{\frac{1+h}{C^2} +i g k}
\label{longF}
\end{equation}
The first term is the Coulomb scattering amplitude and so the second one must be equal to $\exp(2i\sigma_0) (\exp(2i\delta_0)-1)/2ik$, where $\delta_0$ is the phase shift due to the presence of short-range forces. It is connected to the scattering length $\alpha$ and effective range $r_0$ as follows~\cite{31}:
$$
	C^2 \cot \delta_0 + \frac{2}{ka_c} h(ka_c) = -\frac{1}{\alpha k} + \frac{1}{2} r_0 k,
$$
with $h(z) = {\rm Re} \Psi(-i/z) + \ln(z)$ and $\Psi(z)$ is digamma function. For small $z$ we have $h(z) \approx z^2/12$. Investigating near--threshold behavior of $A_0(k)$ we can neglect the effective range term and $h(z)/z$ term and taking into account the fact that $|\delta_0| \ll 1$ rewrite this definition:
\begin{equation}
	C^2 \frac{1}{\delta_0} = -\frac{1}{\alpha k}.
\label{slc}
\end{equation}
Now from equation~(\ref{longF}) it follows
$$
	-\frac{g}{\frac{1+h}{C^2}+igk} = \frac{\exp(2i\delta_0)-1}{2ik},
$$
and expanding left hand side in terms of $k$ and right hand side in terms of $\delta_0$
$$
	-C^2 \frac{g}{1+h} = \frac{\delta_0}{k} = -\alpha C^2.
$$
So finally the desired connection is:
$$
	\frac{g}{1+h} = \alpha.
$$
Corresponding expansion for $f(-k)$ is
$$
	f(-k) = (1+h) \times \left(\frac{1}{C} \exp(-i\sigma_0) + \exp(-i\sigma_0)iC\alpha k \right)
$$
And then finally
\begin{equation}
	|A_0(k)| = {\rm const}\: C \:(1+C^2{\rm Im}\alpha k).
\label{A0C}
\end{equation}

In the case of $p\bar{p}$ system Coulomb force is attractive so $a_c = -\hbar^2/me^2$. The Gamow factor then becomes
$$
	C^2 = \frac{2\pi/k|a_c|}{1-\exp(-2\pi/k|a_c|)}.
$$
For small c.m. momentum $2\pi/k|a_c| \gg 1$ it has the form
$$
	C^2 = \frac{2\pi}{k|a_c|},
$$
and from~(\ref{A0C}) it follows that
\begin{equation}
	|A_0(k)| = {\rm const}\: \sqrt{\frac{2\pi}{k|a_c|}} \:\left(1+\frac{2\pi}{|a_c|}{\rm Im}\alpha\right).
\end{equation}

Comparing~(\ref{A0C}) with~(\ref{A0}) we see that the effect of Coulomb interaction on form factor behaviour becomes negligible as $C^2 \approx 1$, i.e. when $2\pi/k|a_c| \ll 1$. For $p\bar{p}$ system this gives $p \gg 20\:{\rm MeV}/c$.

\section{}
\label{appndxB}
Here we derive the form factor for the following model $N\bar{N}$ interaction. We assume that $N\bar{N}$ potential $V(r)$ can be approximated above some matching distance $R_c$ by an attractive exponential potential $V(r>R_c)\approx-U_{0} \exp(-(r-R_c)/\rho)$ . Below the matching distance $r<R_c$ the $N\bar{N}$ interaction has a deep inner-core such that collision energy can be neglected $V(r) \gg k^2/(2M)$ .

It is possible to obtain the energy dependence of the form factor  under these conditions without any further assumptions.
 For $r<R_c$ the  regular wave-function $\phi(r)$ satisfies boundary conditions:
\begin{equation}
	\phi(0) = 0
\end{equation}
and
\begin{equation}
	\phi '(0) = 1
\end{equation}
and its $k$-dependence can be neglected. It means that at matching distance $R_c$ the   continuity   condition for the logarithmic derivative of the wave-function is also energy independent.  We will parameterize  it in the following way:
 \begin{equation}\label{boundary}
 \phi'(r)/\phi(r)|_{r=R_c}=\frac{z(R_c)}{2\rho} \cot (\Phi_0)
 \end{equation}
 Here $z(R)=2\rho \sqrt{2 M U_0}\exp(-(R-R_c)/(2\rho))$, $M$-is the $N\bar{N}$ reduced mass, $\Phi_0$ is a free  parameter.
For $r>R_c$ the wave function is given by the solution of the Schr\"{o}dinger equation with exponential potential $V(r)=-U_{0} \exp(-(r-R_c)/\rho)$:
\begin{equation}
	\phi(r;k) = B_1(k) J_{2ik\rho}\left(z(r)\right) + B_2(k) J_{-2ik\rho}\left(z(r)\right),
\end{equation}
Using continuity of the wave functions and its derivative at $r=R_c$ we get for the form factor

\begin{equation}
	A_0(k) = {\rm const} \frac{\sinh 2\pi k\rho}{2\pi k\rho} \Gamma(1+2ik\rho) (z_c/2)^{-2ika}\frac{1}{(z_c/2)(J_{-2ik\rho}(z_c) \cos \Phi_0 + J'_{-2ik\rho}(z_c) \sin \Phi_0)}.
\end{equation}
here $z_c \equiv z(R_c)$.
If we additionally assume that $|z_c|\gg 1$, which means that the wave function can be approximated by it's WKB form  near $R_c$ and taking into account the Bessel function large argument  behavior:
 \begin{equation}
	J_{2ik\rho}(2x) \sim \frac{1}{\sqrt{\pi x}} \cos(2x - i\pi k \rho - \frac{\pi}{4}),
\end{equation}
  we  finally get the simplified expression:
 \begin{equation}
	A_0(k)  = {\rm const}\: \frac{\sinh(2\pi k \rho)}{2\pi k \rho} \: \frac{\Gamma(1+2ik\rho) (z_c/2)^{-2ik\rho}} { \cos(\Phi_0 +z_c +i\pi k \rho-\pi/4)},
\end{equation}
and its absolute value:
\begin{equation}
	|A_0(k)| = {\rm const}\sqrt{\frac{\sinh 2\pi k\rho}{2\pi k\rho}} \frac{1}{|\cos(\Phi + i\pi k \rho)|}.
\end{equation}
where $\Phi=\Phi_0 +z_c-\pi/4$.
One can see that the phase $\Phi_0$ has a sense of  phase accumulated in the region $r<R_c$. This explains our choice of the boundary condition parametrization~(\ref{boundary}).
Expanding the expression for the form factor in terms of $k$ we get for the imaginary part of the scattering length:
\begin{equation}
	{\rm Im} \alpha = -\pi \rho \: {\rm Im}\:\tan (\Phi_0),
\end{equation}

\section*{Acknowledgement}
\label{ackn}
We would like to thank J. Carbonell for providing us with the program of computating optical potential.


\end{document}